\documentclass[aps,groupedaddress,superscriptaddress,showpacs]{revtex4-1}
\usepackage{eurosym}
\usepackage{amsmath}
\usepackage{fixmath}
\usepackage{graphicx}
\usepackage{wrapfig}
\usepackage[toc,page]{appendix}
\usepackage{caption}
\usepackage{hyperref}
\usepackage[capitalise]{cleveref}
\usepackage{subcaption}
\usepackage{amsfonts}
\usepackage{xcolor}

\usepackage{dcolumn}
\usepackage{bm}

\begin{document}


\title{Coulomb-induced synchronization of intersubband coherences in highly doped quantum wells and the formation of giant collective resonances}


\author{ Mikhail Tokman}
\affiliation{Department of Electrical and Electronic Engineering, Ariel University, 40700 Ariel, Israel }
\author{ Maria Erukhimova}
\affiliation{Biraghigasse 8, 1130 Vienna, Austria}
\author{Yongrui Wang}
\affiliation{Department of Physics and Astronomy, Texas A\&M University, College Station, TX, 77843 USA}
\author{Alexey Belyanin}
\affiliation{Department of Physics and Astronomy, Texas A\&M University, College Station, TX, 77843 USA}


\date{\today}

\begin{abstract}
Many-body Coulomb interactions drastically modify the optical response of highly doped semiconductor quantum wells leading to a merger of all intersubband transition resonances into one sharp peak at the frequency substantially higher than all single-particle  transition frequencies. Starting from standard density matrix equations for the gas of pairwise interacting fermions within Hartree-Fock approximation, we show that this effect is due to Coulomb-induced synchronization of the oscillations of coherences of all $N$ intersubband transitions and sharp collective increase in their coupling with an external optical field. In the high doping limit,  the dynamics of light-matter interaction is described by the analytic theory of $N$ coupled oscillators which determines new collective normal modes of the system and predicts the frequency and strength of the blueshifted collective resonance. 
\end{abstract}


\maketitle

\section{\label{Intro}Introduction}

Intersubband optical transitions in doped semiconductor quantum wells attracted strong recent interest due to their giant optical nonlinearities, tunability, and ultrafast response,  which promise a broad range of applications from nonlinear frequency mixing, ultrafast saturation, and mode locking to ultra-strong coupling in cavity quantum electrodynamics; see, e.g., recent papers \cite{Mann2021,Nefedkin2021, Piccardo2020, jeannin2021, Kono2019} and references therein.    There has been a number of experimental and theoretical studies of dramatic changes in the intersubband absorption and emission in highly doped semiconductor quantum wells. These changes are the  manifestation of the Coulomb-induced collective effect called the depolarization field \cite{Helm00, Ando82}, when in the presence of the electromagnetic (EM) radiation each electron is affected by an effective field induced by the excitations of other electrons. The main result of such coupling is the macroscopic polarization which is established in the quantum well as a result of collective modes of oscillations of the electron gas. When the electron density is low and only a single subband is occupied, the depolarization field results in a blue shift of the absorption peak with respect to the ``bare'' transition frequency. This resonance frequency corresponds to the so called intersubband plasmon \cite{Wendler93}. The effect of the depolarization field is much more dramatic if the quantum well is highly doped so that several subbands are occupied. In this case, instead of several absorption peaks corresponding to transitions between different subbands, the experiment shows a single strong peak,  blueshifted from all ``bare'' transitions  \cite{Delteil12}. This has been interpreted as the excitation of the collective mode of the system, the so-called multisubband plasmon \cite{Delteil12,Pegolotti14}. The model describing this effect \cite{Pegolotti14,Todorov12} was based on the formalism of ``bosonization'' of the electron gas. In this formalism the Hamiltonian describing the electron gas is reduced to the expression that contains bosonic operators instead of fermionic ones, namely the operators of creation and annihilation of excitations associated with a given intersubband transition. Such bosonic  operators are proportional to the dyadics  ${\hat{\rho }}_{mn}={\hat{a}}^{\dagger }_n{\hat{a}}_m$, where ${\hat{a}}^{\dagger }_n$ and ${\hat{a}}_m$ are the creation and annihilation operators of electrons in corresponding states.  When the populations are assumed constant, the operators  ${\hat{\rho }}_{mn}$ obey the standard bosonic commutation relation $\left[{\hat{\rho }}_{mn}\mathrm{\ },{\hat{\rho }}^{\dagger }_{mn}\right]=\mathrm{const}$. The same operators ${\hat{\rho }}_{mn}$ define the operator of the electric polarization which appears in the total Hamiltonian in the dipole gauge \cite{Babiker83} as an independent variable and describes the effects of dipole-dipole interactions and the coupling of the electronic polarization with a quantized EM field \cite{Todorov12, Tokman13,TokmanAn15,TokmanPRB15}.   

We develop an alternative approach to describe the light-matter dynamics which does not rely on any approximations related to  bosonization and replacement of the fermionic Hamiltonian by effective bosonic one. We obtain the absorption spectrum of the high-density two-dimensional electron gas confined in a quantum well by solving von Neumann density matrix equations taking into account pairwise Coulomb interactions of electrons within the Hartree-Fock (HF) approximation. We show that at high doping the exchange interaction (Fock) terms become insignificant as compared to Hartree terms. Moreover, the exchange interaction effects contributing to the intersubband transition energy renormalization and the coupling of coherences nearly cancel each other. Therefore, the exchange interaction introduces negligible corrections to the spectrum, which is dominated by electron interaction through a common field. Mathematically, the problem is reduced to the system of linearly coupled $2N$ first-order differential equations for coherences excited at the intersubband transitions by an external monochromatic force, where $N$ is a number of the intersubband transitions.  Therefore, the observed spectra can be understood within an intuitive and transparent picture of self-synchronization in a system of $N$ coupled oscillators, which is a universal phenomenon in the nonlinear dynamics, with numerous analogies not only in quantum-well optics (e.g., self-synchronization of quantum-cascade laser modes \cite{wojcik2011} or Coulomb-induced Fermi-edge singularity \cite{kim2013}) and plasma physics (e.g., synchronization of oscillations of free electrons in the collective
field of a Langmuir wave), but across all areas of physics and other sciences \cite{pikovsky2001, SCH07, acebron2005}. We are able to obtain important analytic results, in particular the frequencies and oscillator strengths of the new collective normal modes of the system. The collapse of all transitions into a single absorption peak is naturally explained by the presence of Coulomb-induced couplings between intersubband coherences, similarly to the effect of springs connecting mechanical oscillators. In fact, the mechanical analogy can be made mathematically exact; see Fig.~1 and Appendix C. The action of these ``Coulomb springs'' leads to both the blueshift of the collective resonance frequency and giant enhancement of its oscillator strength.

The paper is organized as follows. In Section~\ref{Hamiltonian} we present the Hamiltonian of the system of Coulomb-coupled identical fermions in a quantum well within the second quantization formalism. In Section~\ref{Dynamics} the von Neumann equations for coherences are derived in Hartree-Fock approximation. The final form of the equations taking into account the Hartree modification of the ground state and Hartree coupling terms, and neglecting the exchange interaction is presented. In Section~\ref{Absorption} the solution of this system of coupled equations for coherences at different intersubband transitions is obtained in a general form. The expression for the absorption spectrum of a highly doped quantum well with several occupied subbands is derived analytically. It represents the superposition of collective oscillation eigenmodes with amplitudes proportional to the oscillator strengths and eigenfrequencies different from the original ``bare'' intersubband transition frequencies.  In subsection~\ref{Coulomb} we prove that the Coulomb interaction leads to a collapse of several excited intersubband transitions into one sharp peak at the frequency substantially higher than all the transition frequencies. This effect is illustrated by the mechanical model of ``Coulomb springs''. The conditions imposed on the electron concentration and the quantum well thickness under which this effect dominates the optical response are formulated in subsection~\ref{Conditions}. The analytic expression for the frequency of a single bright resonance mode is obtained there. In subsection~\ref{Spectra} the numerically calculated absorption spectra of an electron gas with different concentrations and quantum wells with different thicknesses are presented. In Section~\ref{Influence} we investigate analytically and numerically the impact of exchange effects on the absorption spectra and come to the conclusion that they are negligible at high doping. Appendix A describes the eigenfunctions and eigenstates of the Hartree Hamiltonian. Appendix B proves the sum rule for the new collective normal modes of the system. Appendix C derives the equations of motion for the mechanical model of ``Coulomb springs'' and Appendix D evaluates the screening effect coming from higher-order correlations.

\section{\label{Hamiltonian} The model and the Hamiltonian}

In the second quantization form, the Hamiltonian of the system of interacting identical electrons placed in the QW confinement potential and the potential of ions can be written as
\begin{equation} \label{GrindEQ__1_} 
\hat{H}={\hat{H}}_0+{\hat{H}}_{ee}+{\hat{H}}_{ei} 
\end{equation} 
where 
\[{\hat{H}}_0=\int{d^3\bm{r}{\hat{\mathit{\Psi}}}^{\dagger }_e(\bm{r})\left(\frac{{\bm{p}}^2}{2m^*}+V_e(z)\right){\hat{\mathit{\Psi}}}_e(\bm{r})},\] 
is a free-particle term, 
\[{\hat{H}}_{ee}=\frac{1}{2}\int{d^3\bm{r}\int{d^3{\bm{r}}{\bm{'}}{\hat{\mathit{\Psi}}}^{\dagger }_e(\bm{r}){\hat{\mathit{\Psi}}}^{\dagger }_e({\bm{r}}{\bm{'}})V\left(\left|\bm{r}\bm{-}{\bm{r}}{\bm{'}}\right|\right){\hat{\mathit{\Psi}}}_e({\bm{r}}{\bm{'}}){\hat{\mathit{\Psi}}}_e(\bm{r})}},\] 
describes electron-electron interactions, and 
\[{\hat{H}}_{ei}=-N_{2D}\int{d^3\bm{r}\int{d^3{\bm{r}}{\bm{'}}{\hat{\mathit{\Psi}}}^{\dagger }_e(\bm{r}){\hat{\mathit{\Psi}}}_e(\bm{r})V\left(\left|\bm{r}\bm{-}{\bm{r}}{\bm{'}}\right|\right){\rho}_i\left(z{'}\right)}},\] 
describes electron-ion interactions. 
Here $m^*$ is the effective mass,  $V_e(z)$ is the confinement potential, $z$ is the growth direction of the quantum well structure, $V\left(\left|\bm{r}\bm{-}{\bm{r}}{\bm{'}}\right|\right)={e^2}/{{\varepsilon }_0\left|\bm{r}\bm{-}{\bm{r}}{\bm{'}}\right|}$ 
is the Coulomb interaction potential, ${\varepsilon }_0$ is the background dielectric constant, $N_{2D}$ is the sheet doping density, and ${\rho }_i(z)$ is the normalized doping profile of the ions, satisfying $\int{{\rho }_i(z)dz}=1$. 
The operator ${\hat{\mathit{\Psi}}}_e(\bm{r})$ can be expanded using the wave functions which form a complete one-particle basis. The basis functions are not necessarily wave functions which diagonalize a one-particle Hamiltonian 
${\hat{H}}^0=\frac{{\bm{p}}^2}{2m^*}+V_e(z)$; the only requirement is that these are eigenfunctions of the two-dimensional momentum operator
${\bm{p}}_{\bm{\bot }}=-i \frac{\partial }{\partial \bm{R}}$, 
where $\bm{R}$ is the coordinate in the plane of quantum well:
\[{\hat{\mathit{\Psi}}}_e\left(\bm{r}\right)=\sum_{n\bm{k}}{{\varphi }_n(z)\frac{e^{i\bm{kR}}}{\sqrt{S}}}{\hat{a}}_{n\bm{k}},\] 
where ${\hat{a}}_{n\bm{k}}$  is the fermionic annihilation operator in the corresponding state. The quantity $S$ is the normalization area in the QW plane, 
\[\frac{1}{S}\int_S{d^2\bm{R}\int^{\infty }_{-\infty }{dz{\varphi }^*_n(z){\varphi }_m(z)e^{-i\bm{kR}}e^{i{\bm{k}}{\bm{'}}\bm{R}}}}={\delta }_{nm}{\delta }_{\bm{\kappa }\bm{\kappa }\bm{'}}.               \] 
Then we get for the components of the Hamiltonian in Eq.~(\ref{GrindEQ__1_})      
\[{\hat{H}}_0=\sum_{mn}{\sum_{\bm{k}}{H^0_{mn}(\bm{k}){\hat{a}}^{\dagger }_{m\bm{k}}}{\hat{a}}_{n\bm{k}}},\] 

\[{\hat{H}}_{ee}=\frac{1}{2S}\sum_{mnlp}{\sum_{{\bm{k}}_{\bm{1}}{\bm{k}}_{\bm{2}}\bm{q}}{V^{ee}_{mnlp}\left(q\right){\hat{a}}^{\dagger }_{m{\bm{k}}_{\bm{1}}\bm{-}\bm{q}}}{\hat{a}}^{\dagger }_{l{\bm{k}}_{\bm{2}}\bm{+}\bm{q}}{\hat{a}}_{p{\bm{k}}_{\bm{2}}}{\hat{a}}_{n{\bm{k}}_{\bm{1}}}}=\] 
\[=\frac{1}{2S}\sum_{mnlp}{{\left.V^{ee}_{mnlp}\left(q\right)\right|}_{q=0}\sum_{{\bm{k}}_{\bm{1}}{\bm{k}}_{\bm{2}}}{{\hat{a}}^{\dagger }_{m{\bm{k}}_{\bm{1}}}}{\hat{a}}^{\dagger }_{l{\bm{k}}_{\bm{2}}}{\hat{a}}_{p{\bm{k}}_{\bm{2}}}{\hat{a}}_{n{\bm{k}}_{\bm{1}}}}+\] 
\[+\frac{1}{2S}\sum_{mnlp}{\sum^{q\neq 0}_{{\bm{k}}_{\bm{1}}{\bm{k}}_{\bm{2}}\bm{q}}{V^{ee}_{mnlp}(q){\hat{a}}^{\dagger }_{m{\bm{k}}_{\bm{1}}\bm{-}\bm{q}}{\hat{a}}^{\dagger }_{l{\bm{k}}_{\bm{2}}\bm{+}\bm{q}}{\hat{a}}_{p{\bm{k}}_{\bm{2}}}{\hat{a}}_{n{\bm{k}}_{\bm{1}}}}},\]

\[{\hat{H}}_{ei}=-N_{2D}\sum_{mn}{{\left.V^{ei}_{mn}(q)\right|}_{q=0}\sum_{\bm{k}}{{\hat{a}}^{\dagger }_{m\bm{k}}{\hat{a}}_{n\bm{k}}}},      \] 
where
\begin{equation} \label{eq_1star}
V^{ee}_{mnlp}\left(q\right)=\int{dz\int{dz{'}\frac{2\pi e^2}{{\varepsilon }_0q}e^{-q\left|z-z{'}\right|}{\varphi }^*_m(z)}}{\varphi }_n(z){\varphi }^*_l(z{'}){\varphi }_p(z{'})                        
\end{equation}
\[V^{ei}_{mn}\left(q\right)=\int{dz\int{dz{'}\frac{2\pi e^2}{{\varepsilon }_0q}e^{-q\left|z-z{'}\right|}{\varphi }^*_m(z)}{\varphi }_n(z){\rho }_i(z{'})}.     \] 
The charge neutrality condition requires that ${SN}_{2D}=\sum_{n\bm{k}}{n_{n\bm{k}}}$, where $n_{n\bm{k}}=\left\langle {\hat{a}}^{\dagger }_{n\bm{k}}{\hat{a}}_{n\bm{k}}\right\rangle $. The $q=0$ terms should be interpreted as
\[{\left.\frac{2\pi e^2}{{\varepsilon }_0q}e^{-q\left|z-z{'}\right|}\right|}_{q=0}={\mathop{\mathrm{lim}}_{q\to 0} \frac{2\pi e^2}{{\varepsilon }_0q}e^{-q\left|z-z{'}\right|}\ }={\left.V^{2D}(q)\right|}_{q=0}-\frac{2\pi e^2}{{\varepsilon }_0}\left|z-z{'}\right|,   \] 
where $V^{2D}\left(q\right)=\frac{2\pi e^2}{{\varepsilon }_0q}$ is the two-dimensional Fourier transform of the Coulomb potential. Divergence of  $V^{2D}(q)$ can be avoided by considering the screening effect, see Appendix~\ref{app D}.

In the presence of an optical field $\mathcal{E}(t)$ polarized along with the growth direction the Hamiltonian contains another term,
\[{\hat{H}}_{e-ph}=-\mathcal{E}(t)\sum_{mn}{\sum_{\bm{k}}{{\mu }_{mn}{\hat{a}}^{\dagger }_{m\bm{k}}{\hat{a}}_{n\bm{k}}}},      \] 
where ${\mu }_{mn}$ are the dipole matrix elements. When the two indices are equal, ${\mu }_{nn}=ez_{nn}$, where $z_{nn}$ is the average position for level $n$. This element is only relevant for asymmetric QWs, otherwise it is just a constant in the Hamiltonian.


\section{\label{Dynamics}Dynamics in the Hartree basis}

The dynamics of the density matrix elements ${\rho }_{nm}\left(\bm{k}\right)\equiv \left\langle {\hat{a}}^{\dagger }_{m\bm{k}}{\hat{a}}_{n\bm{k}}\right\rangle $ is described by the Heisenberg equations 
\[i\hbar \frac{d}{dt}\left\langle {\hat{a}}^{\dagger }_{m\bm{k}}{\hat{a}}_{n\bm{k}}\right\rangle =\left\langle \left[{\hat{a}}^{\dagger }_{m\bm{k}}{\hat{a}}_{n\bm{k}},\ {\hat{H}}_0+{\hat{H}}_{ee}+{\hat{H}}_{ei}+{\hat{H}}_{e-ph}\right]\right\rangle .    \] 
For commutation with ${\hat{H}}_{ee}$ and  ${\hat{H}}_{ei}$, it can be shown that the terms proportional to  ${\left.V^{2D}(q)\right|}_{q=0}$ give zero. For the rest of the terms we get
\begin{eqnarray}
i\hbar \frac{d}{dt}{\rho }_{nm}\left(\bm{k}\right)=\sum_l{\left({H^0_{nl}\left(\bm{k}\right){\rho }_{lm}\left(\bm{k}\right)-H}^0_{lm}\left(\bm{k}\right){\rho }_{nl}\left(\bm{k}\right)\right)} \nonumber  \\ 
+\frac{1}{S}\sum_{lpg}{{\tilde{V}}^{ee}_{nlgp}{\rho }_{lm}\left(\bm{k}\right)\sum_{{\bm{k}}{\bm{'}}}{{\rho }_{pg}\left({\bm{k}}{\bm{'}}\right)}}-\frac{1}{S}\sum_{lpg}{{\tilde{V}}^{ee}_{lmgp}{\rho }_{nl}\left(\bm{k}\right)\sum_{{\bm{k}}{\bm{'}}}{{\rho }_{pg}\left({\bm{k}}{\bm{'}}\right)}} \nonumber \\  
-N_{2D}\sum_l{\left({\tilde{V}}^{ei}_{nl}{\rho }_{lm}\left(\bm{k}\right)-{\tilde{V}}^{ei}_{lm}{\rho }_{nl}\left(\bm{k}\right)\right)} \nonumber \\ 
-\mathcal{E}(t)\sum_l{\left({\mu }_{nl}{\rho }_{lm}\left(\bm{k}\right)-{\mu }_{lm}{\rho }_{nl}\left(\bm{k}\right)\right)}
\nonumber \\ 
-\frac{1}{S}\sum_{lpg}{\sum_{\bm{q}\bm{\neq }\bm{0}}{V^{ee}_{npgl}\left(q\right){\rho }_{lm}\left(\bm{k}\right){\rho }_{pg}\left(\bm{k}\bm{+}\bm{q}\right)}}+\frac{1}{S}\sum_{lpg}{\sum_{\bm{q}\bm{\neq }\bm{0}}{V^{ee}_{lpgm}(q){\rho }_{nl}\left(\bm{k}\right){\rho }_{pg}\left(\bm{k}\bm{+}\bm{q}\right)}},   
\label{GrindEQ__2_} 
\end{eqnarray} 
where
\begin{equation} \label{GrindEQ__3_} 
{\tilde{V}}^{ee}_{mnlp}=-\int{dz\int{dz{'}\frac{2\pi e^2}{{\varepsilon }_0}\left|z-z{'}\right|{\varphi }^*_m(z)}}{\varphi }_n(z){\varphi }^*_l(z{'}){\varphi }_p(z{'}) 
\end{equation} 
\begin{equation} \label{GrindEQ__4_} 
{\tilde{V}}^{ei}_{mn}=-\int{dz\int{dz{'}\frac{2\pi e^2}{{\varepsilon }_0}\left|z-z{'}\right|{\varphi }^*_m(z)}{\varphi }_n(z){\rho }_i(z{'})}.             
\end{equation} 
In the derivation above, the random phase approximation (RPA) is used, namely we split quadruple correlators and only keep the density matrix elements which are diagonal with respect to $\bm{k}$. 

It is important that the first four lines in Eq.~\eqref{GrindEQ__2_} can be obtained by including only the coupling of electrons  through their collective Coulomb field and their interaction with the optical field, i.e., they follow from the single-particle Hamiltonian including the self-consistent field: 
\[{\hat{H}}^{(1)}={\hat{H}}^0-e\phi \left(z\right)+{\hat{H}}^{(1)}_{e-ph}.        \] 
Here the electric potential $\phi \left(z\right)$ obeys the one-dimensional Poisson's equation
\[{\phi }^{''}_{zz}=-\frac{4\pi }{{\varepsilon }_0}Q\left(z\right),\] 
where the spatial charge density distribution $Q\left(z\right)$ is self-consistently expressed via the density matrix elements:
\[Q\left(z\right)=eN_{2D}{\rho }_i\left(z\right)-e\frac{1}{S}\sum_{mn}{\sum_{\bm{k}}{{\rho }_{mn}\left(\bm{k}\right)}{\varphi }^*_m\left(z\right){\varphi }_n(z)}.      \] 
The last line in Eq.~\eqref{GrindEQ__2_} is due to exchange interaction. It cannot be obtained in the single-particle picture. 

The eigenstates of the single-particle Hamiltonian ${\hat{H}}^H={\hat{H}}^0-e\phi \left(z\right)$, in which the collective field $\phi \left(z\right)$ is self-consistently produced by electrons with an equilibrium diagonal distribution over these particular eigen tates, form the so-called Hartree basis. The Hartree Hamiltonian can be written as 
\begin{eqnarray} \label{GrindEQ__5_} 
{\hat{H}}^H={\hat{H}}^0+\frac{2\pi e^2}{{\varepsilon }_0}N_{2D}\int{dz{'}\left|z-z{'}\right|{\rho }_i\left(z{'}\right)}-\nonumber \\
-\frac{2\pi e^2}{{\varepsilon }_0}\frac{1}{S}\sum_m{\sum_{\bm{k}}{{\rho }_{mm}\left(\bm{k}\right)}\int{dz{'}\left|z-z{'}\right|{\left|{\varphi }_m(z{'})\right|}^2}}. 
\end{eqnarray} 
The equation for eigenfunctions and eigenvalues is
\begin{equation} \label{GrindEQ__6_} 
{\hat{H}}^H{\varphi }_m\left(z\right)\frac{e^{i\bm{kR}}}{\sqrt{S}}=\left(\frac{{\hbar }^2k^2}{2m^*}+E^H_m\right){\varphi }_m\left(z\right)\frac{e^{i\bm{kR}}}{\sqrt{S}}.                 
\end{equation} 
Here ${\rho }_{mm}\left(\bm{k}\right)$ are equilibrium populations obeying the Fermi-Dirac statistics over the self-consistently obtained energies $E^H_m$. The equations that can be used for the numerical calculation of the Hartree basis are presented in the Appendix~\ref{app A}.

Considering the exchange interaction and the interaction with the optical field as perturbations, the equilibrium diagonal distribution over the Hartree states should be used as an unperturbed state of the system. Equation \eqref{GrindEQ__2_} is greatly simplified in the Hartree basis defined by Eqs.~\eqref{GrindEQ__5_}, \eqref{GrindEQ__6_}. In linear approximation with respect to  perturbations, the equations of motion for the nondiagonal density matrix elements  take the form
\begin{eqnarray}
i\hbar \frac{d}{dt}{\rho }_{nm}\left(\bm{k}\right)=\left(E^H_n-E^H_m\right){\rho }_{nm}\left(\bm{k}\right)+\nonumber \\ 
+\frac{1}{S}\sum_{p\neq g}{{\tilde{V}}^{ee}_{nmgp}\sum_{{\bm{k}}{\bm{'}}}{{\rho }_{pg}\left({\bm{k}}{\bm{'}}\right)}}\left({\rho }_{mm}\left(\bm{k}\right)-{\rho }_{nn}\left(\bm{k}\right)\right)+ \nonumber \\ 
 -\frac{1}{S}\sum_{lp}{\sum_{\bm{q}\bm{\neq }\bm{0}}{{\rho }_{pp}\left(\bm{k}\bm{+}\bm{q}\right)\left(V^{ee}_{nppl}\left(q\right){\rho }_{lm}\left(\bm{k}\right)-V^{ee}_{lppm}\left(q\right){\rho }_{nl}\left(\bm{k}\right)\right)}}- \nonumber \\ 
-\frac{1}{S}\sum_{p\neq g}{\sum_{\bm{q}\bm{\neq }\bm{0}}{V^{ee}_{npgm}\left(q\right){\rho }_{pg}\left(\bm{k}\bm{+}\bm{q}\right)}}\left({\rho }_{mm}\left(\bm{k}\right)-{\rho }_{nn}\left(\bm{k}\right)\right)- \nonumber \\ 
\label{GrindEQ__7_} 
-\mathcal{E}(t){\mu }_{nm}\left({\rho }_{mm}\left(\bm{k}\right)-{\rho }_{nn}\left(\bm{k}\right)\right) .               
\end{eqnarray}

If the exchange interaction is neglected, i.e., the third and the fourth rows in Eq.~\eqref{GrindEQ__7_} are dropped, one can sum over $\bm{k}$ and obtain much more compact equations for the dynamics of variables ${\rho }_{nm}=\frac{1}{S}\sum_{\bm{k}}{{\rho }_{nm}\left(\bm{k}\right)}$:
\begin{equation} \label{GrindEQ__8_}
i\hbar \frac{d}{dt}{\rho }_{nm}=\left(E^H_n-E^H_m\right){\rho }_{nm}+\sum_{p\neq g}{{\tilde{V}}^{ee}_{nmgp}}\left({\rho }_{mm}-{\rho }_{nn}\right){\rho }_{pg}-\mathcal{E}\left(t\right){\mu }_{nm}\left({\rho }_{mm}-{\rho }_{nn}\right)-i\hbar \mathrm{\Gamma }{\rho }_{nm}. 
\end{equation}
Here we have added the relaxation term in its simplest form. Equation \eqref{GrindEQ__8_}  clearly demonstrates that the Coulomb interaction creates linear coupling of effective electron oscillators at different transitions between the subbands dressed by the self-consistent field. This coupling is stronger with increasing population differences. Note that Eq.~\eqref{GrindEQ__8_} contains total populations due to summation over $\bm{k}$ and therefore the population differences can be large despite the Pauli blocking of some $k$-states.  

In what follows, we analyze the absorption of a highly-doped QW system. We use Eq.~\eqref{GrindEQ__8_}, where exchange terms are dropped. In Section~\ref{Influence} we evaluate the effect of exchange terms on the absorption spectrum.


\section{\label{Absorption}The absorption spectrum of highly doped quantum wells}

Equations \eqref{GrindEQ__8_} represent a system of first-order  differential equations for $N_l$  linearly coupled variables ${\rho }_j={\rho }_{nm}$ in the presence of an ``external force''. Here $N_l$ is a number of discrete levels (Coulomb-dressed subbands) involved in the interaction. The corresponding number of the transitions is $N_t\mathrm{=}\frac{1}{2}N_l\left(N_l-1\right)$. Introducing the index numerating the transitions $j=\left\{nm\right\}$, where the transitions $\left\{nm\right\}=j$ and $\left\{mn\right\}=j{'}$ are counted separately, we can rewrite Eqs.~\eqref{GrindEQ__8_} in the form 
\begin{equation} \label{GrindEQ__9_} 
{\dot{\rho }}_j=-i\sum^{2N_t}_{l=1}{Z_{jl}{\rho }_l}+if_j\left(t\right)-\mathrm{\ }\mathrm{\Gamma }{\rho }_j. 
\end{equation} 
Here the elements of matrix $\mathbb{Z}$ are given by 
\begin{equation} \label{GrindEQ__10_} 
{\mathrm{Z}}_{jl}={{\omega }_j\delta }_{jl}+ \frac{e^2}{\hbar }I_{jl{'}}\mathrm{\Delta }n_j, 
\end{equation} 
where the notations are 
\begin{equation} \label{GrindEQ__11_} 
{\omega }_j=\frac{1}{\hbar }\ \left(E^H_n-E^H_m\right), I_{jl{'}}=\frac{1}{e^2}{\tilde{V}}^{ee}_{nmgp},\  \mathrm{\Delta }n_j={\rho }_{mm}-{\rho }_{nn},\  f_j\left(t\right)=\frac{1}{\hbar }{\mu }_j\mathcal{E}\left(t\right)\mathrm{\Delta }n_j, {\mu }_j={\mu }_{nm} 
\end{equation} 
for $j=\left\{nm\right\},\ \ l=\{pg\}$, $\ \ l{'}=\{gp\}$.

Considering a monochromatic external field,
\[\mathcal{E}\left(t\right)=Re(E^{\omega }e^{-i\omega t}),         \] 
we are looking for the induced solution of Eq.~\eqref{GrindEQ__9_}, 
\[{\rho }_j={\rho }^{\omega }_je^{-i\omega t}+{\rho }^{-\omega }_je^{i\omega t},         \] 
for which the differential equations Eq.~\eqref{GrindEQ__9_} are reduced to the algebraic ones: 
\begin{equation} \label{GrindEQ__12_} 
-i\omega {\rho }^{\omega }_j=-i\sum^{2N_t}_{l=1}{Z_{jl}{\rho }^{\omega }_l}+if^{\omega }_j-\mathrm{\Gamma }{\rho }^{\omega }_j,             
\end{equation} 
where
\begin{equation} \label{GrindEQ__13_} 
f^{\omega }_j=\frac{1}{2\hbar }{\mu }_jE^{\omega }\mathrm{\Delta }n_j.                
\end{equation} 
Equation \eqref{GrindEQ__12_} can be presented in the vector form, where the dimension of vector space is equal to  $2N_t$:
\[-i\omega {\bm{\rho }}^{\bm{\omega }}=-i\mathbb{Z}{\bm{\rho }}^{\bm{\omega }}+i{\bm{f}}^{\bm{\omega }}-\mathrm{\Gamma }{\bm{\rho }}^{\bm{\omega }}.        \] 

The averaged dipole moment per unit area of the quantum well, excited as a response to the incident EM wave is calculated as 
\[P(t)=\sum_{m,n}{{\mu }_{mn}{\rho }_{nm}}=\sum^{2N_t}_{j=1}{{\mu }_j\left({\rho }^{\omega }_je^{-i\omega t}+{\rho }^{-\omega }_je^{i\omega t}\right)}. \] 
For simplicity, we can assume that the matrix elements of the dipole moment are real, so that ${\mu }_{mn}={\mu }_{nm}$. Effective absorbance, which determines the energy absorbed per unit area of the layer, is given by 
\begin{equation} \label{GrindEQ__14_} 
\mathrm{\Sigma }\left(\omega \right)=\frac{4\omega}{c} {\rm Im} \left(\sum^{2N_t}_{j=1}{{\mu }_j{\rho }^{\omega }_j/E^{\omega }} \right).  \end{equation} 
If the Coulomb interaction of electrons, and therefore the coupling of oscillations at different transitions, is neglected, the matrix $\mathbb{Z}$  in Eq.~\eqref{GrindEQ__10_} is  diagonal, ${\mathrm{Z}}_{jl}={\omega }_j{\delta }_{jl}$, and the induced solution at frequency $\omega $ is defined by the standard Lorentzian:
\begin{equation} \label{GrindEQ__15_} 
{\rho }^{\omega }_j=\frac{if^{\omega }_j}{i\left({\omega }_j-\omega \right)+\mathrm{\Gamma }}.                           
\end{equation} 
In this case the absorbance given by Eq.~\eqref{GrindEQ__14_} yields
\begin{equation} \label{GrindEQ__16_} 
\mathrm{\Sigma }(\omega )= \frac{4 \omega}{c} {\rm Re}  \left(\sum^{2N_t}_{j=1}{\frac{\mu_j f^{\omega }_j/E^{\omega }}{ i\left({\omega }_j-\omega \right)+\mathrm{\Gamma }}}\right).                              
\end{equation} 
Taking into account that $\omega_j=-\omega_{j{'}}$,  $\mu_j=\mu_{j{'}}$,  $f^{\omega }_j=-f^{\omega }_{j{'}}$, Eq.~\eqref{GrindEQ__16_} can be transformed into 
\begin{equation} \label{GrindEQ__17_} 
\mathrm{\Sigma }(\omega )=\frac{e^2N_{2D}}{cm^*}Re\left(\sum^{N_t}_{j({\omega }_j>0)}{F_j\frac{4\omega }{2\mathrm{\Gamma }\omega +i\left({{\omega }_j}^2-{\omega }^2\right)}}\right).           
\end{equation} 
Here the dimensionless parameter $F_j$ is the ``oscillation strength'' of the transition with frequency ${\omega }_j$, multiplied (as compared with the standard definition) by the population difference at this transition normalized to the sheet doping density $\mathrm{\Delta }n_j/N_{2D}$:
\begin{equation} \label{GrindEQ__18_} 
F_j=\frac{2m^*}{e^2N_{2D}}{\omega }_j{\mu }_jf^{\omega }_j/E^{\omega }=\frac{m^*{\mu }^2_j{\omega }_j}{\hbar e^2}\frac{\mathrm{\Delta }n_j}{N_{2D}}.            
\end{equation} 
These modified ``oscillation strengths'' still obey the sum rule,
\[\sum^{2N_t}_{j=1}{F_j}=1.  \] 
This can be proven with the use of the relation   $\sum_m{{\left|{\mu }_{mn}\right|}^2{\omega }_{mn}}=\frac{e^2\hbar }{2m}$, which is true for any one-dimensional Hamiltonian. 
If the transitions are well resolved, the absorption spectrum Eq.~\eqref{GrindEQ__17_} represents the combination of resonant lines with the peak absorbance values given by 
\[\mathrm{\Sigma }(\omega_j)\approx \frac{2e^2N_{2D}}{cm^*}\frac{F_j}{\mathrm{\Gamma }}.        \]

The Coulomb interaction leads to coupling of oscillations at different transitions and enables a dramatic modification of the absorption spectra. The matrix $\mathbb{Z}$ in Eq.~\eqref{GrindEQ__10_} acquires off-diagonal elements.   By the linear change of variables
\[{\tilde{\rho }}_l=B_{lj}{\rho }_j\] 
the matrix $\mathbb{Z}$ can be transformed to the diagonal form, so that for the new variables ${\tilde{\rho }}_l$  Eq.~\eqref{GrindEQ__12_} takes the form 
\begin{equation} \label{GrindEQ__19_} 
-i\omega {\tilde{\rho }}^{\omega }_l=-i{{\mathrm{\Omega }}_l\tilde{\rho }}^{\omega }_l+i{\tilde{f}}^{\omega }_l-\mathrm{\Gamma }{\tilde{\rho }}^{\omega }_l.              
\end{equation} 
The transformation matrix $\mathcal{B}=\left\{B_{lj}\right\}$   is composed of eigenvectors of the transposed Coulomb coupling matrix ${\mathbb{Z}}^T$:
\[\sum_i{Z_{ij}B_{li}}={\mathrm{\Omega }}_lB_{lj},         \] 
or in equivalent form, ${\mathbb{Z}}^T{\bm{B}}_{\bm{l}}={\mathrm{\Omega }}_l{\bm{B}}_{\bm{l}}$. The eigenvalues ${\mathrm{\Omega }}_l$ of matrix $\mathbb{Z}$ (or matrix ${\mathbb{Z}}^T$) are the frequencies of eigenmodes in the system of coupled oscillators. The ``force vector'' ${\bm{f}}^{\bm{\omega }}$ is transformed by the same matrix as 
\[{\tilde{f}}^{\omega }_l=B_{lj}f^{\omega }_j.          \] 
The new components of the ``force vector'' are the projections of this vector onto the directions defined by the vectors ${\bm{B}}_{\bm{l}}$\textbf{ }, i.e., they can be calculated as a scalar product
\begin{equation} \label{GrindEQ__20_} 
{\tilde{f}}^{\omega }_l=\left({\bm{B}}_{\bm{l}}\cdot \bm{f}^{\bm{\omega }}\right).                
\end{equation} 
The solution of Eq.~\eqref{GrindEQ__19_} has a simple form, similar to Eq.~\eqref{GrindEQ__15_}:
\begin{equation} \label{GrindEQ__21_} 
{\tilde{\rho }}^{\omega }_l=\frac{i{\tilde{f}}^{\omega }_l}{i\left({\mathrm{\Omega }}_l-\omega \right)+\mathrm{\Gamma }}.                
\end{equation}

Applying the inverse transformation to Eq.~\eqref{GrindEQ__21_} and substituting the result into Eq.~\eqref{GrindEQ__14_} we get the following expression for the absorbance:
\begin{equation} \label{GrindEQ__22_} 
\mathrm{\Sigma }(\omega )= \frac{4\omega}{c} {\rm Re}\left(\sum^{{2N}_t}_l{\frac{ {\tilde{\mu }}_l{\tilde{f}}^{\omega }_l/E^{\omega }}{i\left({\mathrm{\Omega }}_l-\omega \right)+\mathrm{\Gamma }}}\right) .              
\end{equation} 
It looks exactly like Eq.\eqref{GrindEQ__16_}, but with different resonance frequencies, dipole moments, and external forces, defined for the new collective normal modes of the system. Note that the effective ``dipole vector'' $\bm{\mu }$ is transformed according to the operator which differs from $\mathcal{B}$:
\[{\tilde{\mu }}_l={{\left({\mathcal{B}}^{-1}\right)}^H}_{lj}{\mu }_j.         \] 
It can be shown that new components of the ``dipole vector'' are the projections of this vector onto the directions defined by the eigenvectors of matrix $\mathbb{Z}$ ($\mathbb{Z}{\bm{D}}_{\bm{l}}={\mathrm{\Omega }}_l{\bm{D}}_{\bm{l}}\bm{)}$:
\begin{equation} \label{GrindEQ__23_} 
{\tilde{\mu }}_l=\left({\bm{D}}_{\bm{l}}\bm{\cdot }\bm{\mu }\right).                
\end{equation} 
The matrix $\mathbb{Z}\mathrm{\ }$ is not symmetric, and therefore its eigenvectors are not orthonormal, and the matrix $\mathcal{B}$ is not unitary. However, since matrix $\mathbb{Z}$ obeys the following relation, 
\[Z_{ij}=-Z_{j{'}i{'}},\] 
for every number $l$ which counts a new normal mode, there exists such a number $l{'}$ that ${\mathrm{\Omega }}_l={-\mathrm{\Omega }}_{l{'}}$, $B_{lj}=B_{l{'}j{'}}$, $D_{lj}=D_{l{'}j{'}}$. As result, we have ${\tilde{\mu }}_l={\tilde{\mu }}_{l{'}}$ , ${\tilde{f}}_l={-\tilde{f}}_{l{'}}$. Thus we can rewrite the expression for the absorption spectrum Eq.~\eqref{GrindEQ__22_} in the form  similar to Eq.~\eqref{GrindEQ__17_},
\begin{equation} \label{GrindEQ__24_} 
\mathrm{\Sigma }(\omega )=\frac{e^2N_{2D}}{c m^*}Re\left(\sum^{N_t}_{l({\mathrm{\Omega }}_l>0)}{{\tilde{F}}_l\frac{4\omega }{2\mathrm{\Gamma }\omega +i\left({\mathrm{\Omega }}^2_l-{\omega }^2\right)}}\right).           
\end{equation} 
Here the ``oscillator strengths'' of new normal oscillators are introduced: 
\begin{equation} \label{GrindEQ__25_} 
{\tilde{F}}_l=\frac{2m^*}{e^2N_{2D}}{{\mathrm{\Omega }}_l{\tilde{\mu }}_l{\tilde{f}}^{\omega }_l}/{E^{\omega }}=\frac{2m^*}{e^2N_{2D}}{\mathrm{\Omega }}_l\left({\bm{D}}_{\bm{l}}\bm{\cdot }\bm{\mu }\right)\left({\bm{B}}_{\bm{l}}\bm{\cdot }{\bm{f}}^{\bm{\omega }}\right)/E^{\omega }.          
\end{equation} 
The absorption spectrum in Eq.~\eqref{GrindEQ__24_} represents the superposition of resonant lines at frequencies of the new normal modes ${\mathrm{\Omega }}_l$ with the peak absorbance values proportional to new ``oscillator strengths'' ${\tilde{F}}_l$: 
\[\mathrm{\Sigma }(\omega ={\mathrm{\Omega }}_l)\approx \frac{2e^2N_{2D}}{cm^*}\frac{{\tilde{F}}_l}{\mathrm{\Gamma }}.         \] 
It is remarkable that the sum rule holds true for the new ``oscillation strengths'' as well (see the Appendix~\ref{app B} for the proof):
\begin{equation} \label{GrindEQ__26_} 
\sum^{2N_t}_{l=1}{{\tilde{F}}_l}=1.                
\end{equation}

To summarize this section, the Coulomb coupling of coherences at different intersubband transitions leads to the shift of resonant frequencies and redistribution of the ``oscillation strengths'' between new normal modes. In the next section we show that such redistribution leads to a dramatic effect in which most of the absorbance occurs  at one of the normal mode frequencies, which is  strongly blueshifted with respect to all ``bare'' intersubband transition frequencies.


\subsection{\label{Coulomb}Coulomb-induced self-synchronization of dipole oscillations. The``Coulomb springs'' regime}

In this Section investigate the properties of the Coulomb coupling matrix $\mathbb{Z}$ (Eq.~\eqref{GrindEQ__10_}) and show that under certain conditions the eigenvector of this matrix corresponding to one of the normal modes is optimally oriented with respect to the ``force vector'', so that  the ``oscillation strength'' for this normal mode dominates and reaches its maximum value (${\tilde{F}}_m={\tilde{F}}_{m{'}}\approx \frac{1}{2}$). In other words, oscillations at different intersubband transitions get self-synchronized to produce one powerful collective mode of oscillations. We also show that the eigenfrequency for this mode is large compared with ``bare'' frequencies of intersubband transitions.  

We assume that the doping is high enough, so that  several subbands are populated in equilibrium, and there are $2N_t$ intersubband  transitions with significant dipole moments and total population difference. As an estimation, for an isolated symmetric QW $N_t$ is a number of populated subbands and the relevant transitions are those between neighboring subbands, because they tend to have a much larger transition dipole matrix element as compared to the transitions between more distant subbands. One can of course modify and control the transition dipole moments and frequencies on demand by designing asymmetric coupled QW structures. 

For analytic illustration of the effect, we take some averaged values of the transition frequencies, overlap integrals, dipole moments, and population differences for all transitions:
\begin{equation} \label{GrindEQ__27_} 
\left|{\omega }_j\right|\approx {\omega }_0, \left|\frac{e^2}{\hbar }I_{jl}\mathrm{\Delta }n_j\right|\approx \mathrm{\Omega }, \left|{\mu }_j\right|\approx \mu ,  \left|f^{\omega }_j\right|=\left|\frac{1}{2\hbar }{\mu }_jE^{\omega }\mathrm{\Delta }n_j\right|\approx f         
\end{equation} 
 for all $j$. In a real system these parameters for different transitions are different, and the resulting response will differ from the ideal one, but the basic reasoning remains the same.

Since there is a correspondence between the overlap integrals $I_{ij}$ (see Eqs.~\eqref{GrindEQ__11_} and \eqref{GrindEQ__3_}) and the dipole moments,
\[{\rm sign}\left(I_{ij}\right)={\rm sign}\left({\mu }_i{\mu }_j\right),        \] 
the coupling matrix $\mathbb{Z}$ can be presented in the following form:
\begin{equation} \label{GrindEQ__28_} 
\mathbb{Z}\approx {\omega }_0\left( \begin{array}{cc}
\hat{1} & 0 \\ 
0 & -\hat{1} \end{array}
\right)+\mathrm{\Omega }\left( \begin{array}{cc}
\mathbb{Q} & \mathbb{Q} \\ 
\mathrm{-}\mathbb{Q} & \mathrm{-}\mathbb{Q} \end{array}
\right)=\left( \begin{array}{cc}
{\mathbb{Z}}_+ & {\mathbb{Z}}_{\pm } \\ 
{\mathbb{Z}}_{\mp } & {\mathbb{Z}}_- \end{array}
\right),                      
\end{equation} 
where  
\[{\mathrm{dim} \left(\hat{1}\right)\ }={\mathrm{dim} \left(\mathbb{Q}\right)\ }=N_t\times N_t,    \] 
\[Q_{ij}={\delta }_i{\delta }_j, {\delta }_i= {\rm sign}\left({\mu }_i\right).\] 
Here the numbering order of the subbands is chosen in such a way that ${\omega }_j>0$ for $1\le j\le N_t$ and  $j{'}=j+N_t$. Then the ``dipole vector'' and ``force vector'' are equal to
\begin{equation} \label{GrindEQ__29_} 
\bm{\mu }=\mu \left( \begin{array}{c}
\bm{\delta } \\ 
\bm{\delta } \end{array}
\right)=\left( \begin{array}{c}
{\bm{\mu }}_{\bm{+}} \\ 
{\bm{\mu }}_{\bm{-}} \end{array}
\right),\ \ \ \ \ \ \ \bm{f}=f\left( \begin{array}{c}
\bm{\delta } \\ 
\bm{-}\bm{\delta } \end{array}
\right)=\left( \begin{array}{c}
{\bm{f}}_{\bm{+}} \\ 
{\bm{f}}_{\bm{-}} \end{array}
\right),\  
\end{equation} 
where ${\mathrm{dim} \left(\bm{\delta }\right)\ }=N_t$. It is taken into account in Eq.~\eqref{GrindEQ__28_} and Eq.~\eqref{GrindEQ__29_} that ${\mu }_j={\mu }_{j{'}}$, ${\mathrm{\Delta }n}_j={\mathrm{-}\mathrm{\Delta }n}_{j{'}}$, and ${\omega }_j=-{\omega }_{j{'}}$.

Consider the matrix  ${\mathbb{Z}}_+$, which is the ``positive frequency'' part of matrix $\mathbb{Z}$. The eigenvector of  ${\mathbb{Z}}_+\ $ which corresponds to the in-phase addition of all oscillators, and therefore to the maximum  increase of the corresponding eigenvalue, turns out to be ``co-directional'' with both the ``dipole vector'' ${\bm{\mu }}_{\bm{+}}$ and the ``force vector'' ${\bm{f}}_{\bm{+}}$, so that the corresponding scalar products are maximal.  Such an eigenvector is proportional to $\bm{\delta }$. Taking into account the coupling with ``negative frequency'' vector components, we search for the optimal eigenvector of matrix $\mathbb{Z}$ in the following form (marked by index ``m''):
\[{\bm{D}}_{\bm{m}}=C\left( \begin{array}{c}
\bm{\delta } \\ 
(\alpha -1)\bm{\delta } \end{array}
\right),         \] 
where $C$ is the normalization factor, whereas the parameter $\alpha $ is found from the equation 
\[\mathbb{Z}{\bm{D}}_{\bm{m}}=C\mathrm{\Omega }\left( \begin{array}{c}
\left(\frac{{\omega }_0}{\mathrm{\Omega }}+\alpha N_t\right)\bm{\delta } \\ 
\left(\frac{{\omega }_0}{\mathrm{\Omega }}-\alpha \frac{{\omega }_0}{\mathrm{\Omega }}-\alpha N_t\right)\bm{\delta } \end{array}
\right)={\mathrm{\Omega }}_mC\left( \begin{array}{c}
\bm{\delta } \\ 
(\alpha -1)\bm{\delta } \end{array}
\right).    \] 
Solving for it, we find 
\[\alpha =-\frac{{\omega }_0}{N_t\mathrm{\Omega }}\pm \sqrt{{\left(\frac{{\omega }_0}{N_t\mathrm{\Omega }}\right)}^2+\frac{{2\omega }_0}{N_t\mathrm{\Omega }}}.        \] 
The frequency of the normal mode which corresponds to this eigenvector is
\[{\mathrm{\Omega }}_m=\pm \sqrt{{{\omega }_0}^2+{2\omega }_0N_t\mathrm{\Omega }}.        \] 
Under the condition
\begin{equation} \label{GrindEQ__30_} 
N_t\mathrm{\Omega }\gg {\omega }_0 
\end{equation} 
the frequency of this normal mode turns out to be much larger than a typical bare transition frequency
\begin{equation} \label{GrindEQ__31_} 
{\mathrm{\Omega }}_m\approx \sqrt{2{\omega }_0N_t\mathrm{\Omega }}\gg {\omega }_0.               
\end{equation} 
For the effective dipole moment corresponding to this normal oscillator and for the ``force vector'' component, from Eq.~\eqref{GrindEQ__20_} and Eq.~\eqref{GrindEQ__23_} we obtain 
\[{\tilde{\mu }}_m\approx \mu {\left({{\omega }_0N_t}/{2\mathrm{\Omega }}\right)}^{1/4}, {\tilde{f}}_m\approx f{\left({{\omega }_0N_t}/{2\mathrm{\Omega }}\right)}^{1/4} .     \] 
As a result, the ``oscillator strength'', defined by Eq.\eqref{GrindEQ__25_}, is 
\[{\tilde{F}}_m=\frac{2m^*}{e^2N_{2D}}{\mathrm{\Omega }}_m{\tilde{\mu }}_m{\tilde{f}}^{\omega }_m/E^{\omega }\approx \frac{2m^*}{e^2N_{2D}}N_t{\omega }_0\mu f/E^{\omega }_z.     \] 

It shows that the ``oscillator strength'' at the frequency ${\mathrm{\Omega }}_m$ is $N_t$ times higher as compared with the oscillator strengths at uncoupled transitions Eq.~\eqref{GrindEQ__18_}. By virtue of the sum rule for the new ``oscillation strengths'' Eq.~\eqref{GrindEQ__26_}, an almost total suppression of the optical response at all other normal frequencies will take place. From Eq.~\eqref{GrindEQ__26_} we obtain that ${\tilde{F}}_m=1/2.$  Furthermore, according to Eq.~\eqref{GrindEQ__31_} the frequency of this bright mode is $\sqrt{{2N_t\mathrm{\Omega }}/{{\omega }_0}}$ times higher than bare transition frequencies. Note that even if one subband were populated, $N_t=1$, the frequency of the effective oscillator taking into account Coulomb interaction of electrons would be still $\sqrt{1+2\Omega/\omega_0}$ larger than the ``bare'' transition frequency. If several subbands are populated the frequency of the bright mode gets even higher.

Under the ideal conditions when the whole ``oscillation strength'' is concentrated in one oscillator the absorbance at this frequency reaches the value
\begin{equation} \label{GrindEQ__32_} 
{\mathrm{\Sigma }}_{max}\mathrm{=}\mathrm{\Sigma }(\omega ={\mathrm{\Omega }}_m)\approx \frac{e^2N_{2D}}{cm^*\mathrm{\Gamma }}.              
\end{equation} 


\begin{figure}[b]
\center{\includegraphics[width=0.6\textwidth]{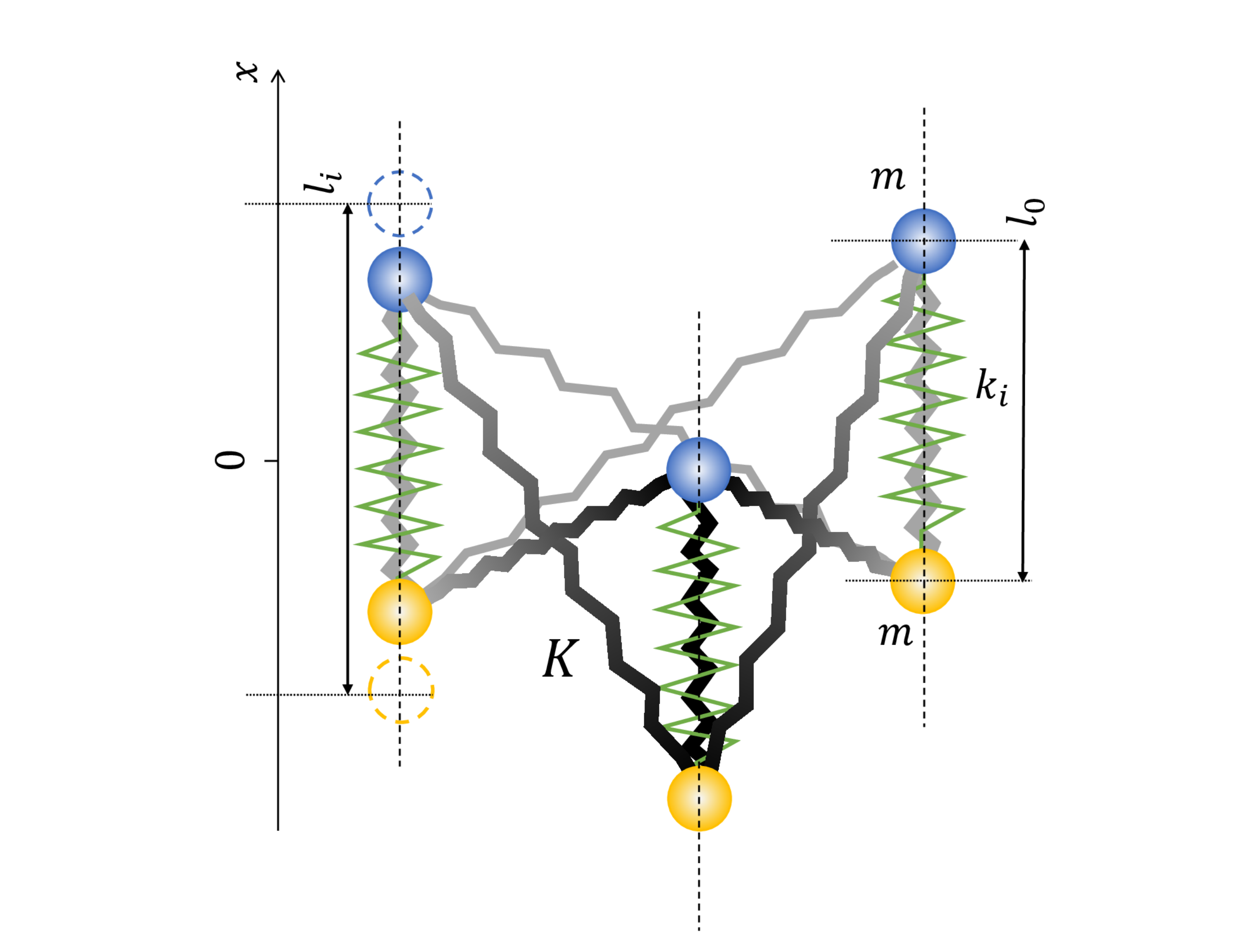}}
\caption{\label{Fig1} Mechanical model of the ``Coulomb springs'' effect in a high-density regime. The oscillations of dipole moments at populated intersubband transitions (here the number of active transitions $N_t=3$) are modeled by vertical vibrations of masses $m$ on green springs with corresponding spring constants $k_j$. The allowed motion is one dimensional (along $x$ axis), and only the oscillations with immobile center of mass of each oscillator are considered. The Coulomb coupling is modeled by grey springs which tie each ``upper'' mass with all ``lower'' masses and vice versa. These springs are characterized by spring constant $K$ which increases with increasing electron density in a QW. Each oscillator can be independently excited by an external force. Under the condition  $KN_t\gg k_j$ the collective in-phase oscillation with frequency defined by the spring constant of the ``Coulomb spring'' and proportional to $\sqrt{N_t}$ is excited most efficiently; see the solution in Appendix~\ref{app C}. }
\end{figure}  


To summarize, Coulomb interaction leads to effective synchronization of the oscillations of coherences at different intersubband transitions. This effect can be illustrated by a simple mechanical model of coupled oscillators as in the sketch shown in Fig.~\ref{Fig1}, where corresponding equations are in Appendix~\ref{app C}. Each active intersubband transition can be modeled by a classical oscillator (the masses on a green spring) with frequency ${\omega }_j$.   The Coulomb couplings of oscillators are shown by the effective additional grey ``springs''. These ``Coulomb springs''  synchronize the  oscillations in phase for all oscillators, which also leads to an increase of eigenfrequency.


\subsection{\label{Conditions}Conditions for a strong Coulomb effect}

Let us analyze the conditions for strong modification of the absorption spectrum in a QW with thickness $L$ and two-dimensional electron gas density  $N_{2D}$. These two parameters determine the spectrum for a given shape of the quantum well potential and material parameters.

The parameter $\mathrm{\Omega }$ is the characteristic frequency, which is a measure of the influence of Coulomb effects on the oscillations of the dipole moment in a QW. Its magnitude scales as (see Eq.~\eqref{GrindEQ__27_})
\begin{equation} \label{GrindEQ__33_} 
\Omega \sim \frac{e^2}{\hbar }\left\langle I_{jl}\right\rangle \left\langle \Delta n_j\right\rangle .                
\end{equation} 
For strong Fermi degeneracy the population of the $n$th subband is equal to ${\rho }_{nn}=\frac{m^*}{\pi {\hbar }^2}\left(E_F-E_n\right)$, where $E_F$ is the Fermi energy. Hence it follows that if several subbands are populated, i.e. $N_t>1$, the averaged population difference at the transitions between neighboring levels a can be estimated as 
\begin{equation} \label{GrindEQ__35_} 
\left\langle \mathrm{\Delta }n_j\right\rangle \mathrm{\sim } \frac{m^*}{\pi \hbar }\left\langle {\omega }_j\right\rangle ,\ \ \ \ \ \left\langle {\omega }_j\right\rangle ={\omega }_0 .             
\end{equation} 
More than one subband is populated if $N_{2D}>\frac{m^*}{\pi \hbar }\omega_1 $, 
where ${\omega }_1$ is the transition frequency between the first two levels.

The magnitude of one-dimensional overlap integrals $I_{jl}$ is proportional to the QW thickness and can be estimated as
\begin{equation} \label{GrindEQ__36_}
\left\langle I_{jl}\right\rangle \sim \frac{\pi }{\varepsilon_0} L J,               
\end{equation}
where $J$ is the dimensionless factor defined by the shape of a QW potential. This parameter does not change much for different transitions between neighboring levels, For example, in a square potential $J\sim 0.2$. 

It follows from Eqs.~\eqref{GrindEQ__33_}, \eqref{GrindEQ__35_}, and \eqref{GrindEQ__36_} that 
\[  \Omega \sim \omega_0\frac{L}{\alpha^*}\frac{J}{\varepsilon_0}, \] 
where ${\alpha }^*=\frac{{\hbar }^2}{e^2m^*}$ is an analog of the Bohr's radius defined for an effective electron mass $m^*$. For example, for GaAs QWs with $m^*=0.067m_e$ the value of $\alpha^*\approx 0.8$ nm. The condition Eq.~\eqref{GrindEQ__30_} of strong Coulomb modification of the absorption spectrum in a QW with several populated subbands can now be written as 
\begin{equation} \label{GrindEQ__37_} 
N_t\frac{L}{{\alpha }^*}\frac{J}{{\varepsilon }_0}\gg 1.                
\end{equation} 
It means that the quantum well thickness multiplied by the number of populated subbands must be sufficiently large. Taking the above parameters and  $\varepsilon_0 \sim 13$ for the background dielectric constant \cite{Li99}, the latter condition becomes 
\begin{equation} \label{GrindEQ__38_} 
N_t L \gg 50\;  {\rm  nm}.  
\end{equation} 

 As follows from Eq.~\eqref{GrindEQ__31_} the resonance frequency of the main peak under the condition \eqref{GrindEQ__37_} becomes 
\begin{equation} \label{GrindEQ__39_} 
{\mathrm{\Omega }}_m\sim {\omega }_0\sqrt{2N_t\frac{L}{{\alpha }^*}\frac{J}{{\varepsilon }_0}}.               
\end{equation} 
The intersubband transition frequencies ${\omega }_0\ $ also depend on the QW thickness. The typical scaling is 
\begin{equation} \label{GrindEQ__40_} 
{\omega }_0\sim {\omega }^*{\left(\frac{{\alpha }^*}{L}\right)}^2.                
\end{equation} 
Here  ${\omega }^*=\frac{e^4m^*}{{\hbar }^3}\pi D$, where $D$ is a dimensionless factor between 1 and 10 which is defined by the quantum well potential shape distorted in some way due to the Hartree effect. For example, in an infinite square potential and neglecting the Hartree contribution, we have $D=\frac{3}{2}\pi $. 

Taking into account the dependence of transition frequencies on the quantum well thickness, Eq.~\eqref{GrindEQ__40_}, the condition for several subbands to be occupied ($N_t>1$)  is reduced to 
\[N_{2D}L^2>D,\] 
so that for QW thicknesses $L\sim 10-20$ nm, the 2D electron density needed to populate several subbands is of the order of $N_{2D}\sim 10^{12}-10^{13}$ cm$^{-2}$, as in the experiments \cite{Delteil12,Delteil13}.

\subsection{\label{Spectra} Examples of absorption spectra}

\begin{figure}
\center{\includegraphics[width=0.7\textwidth]{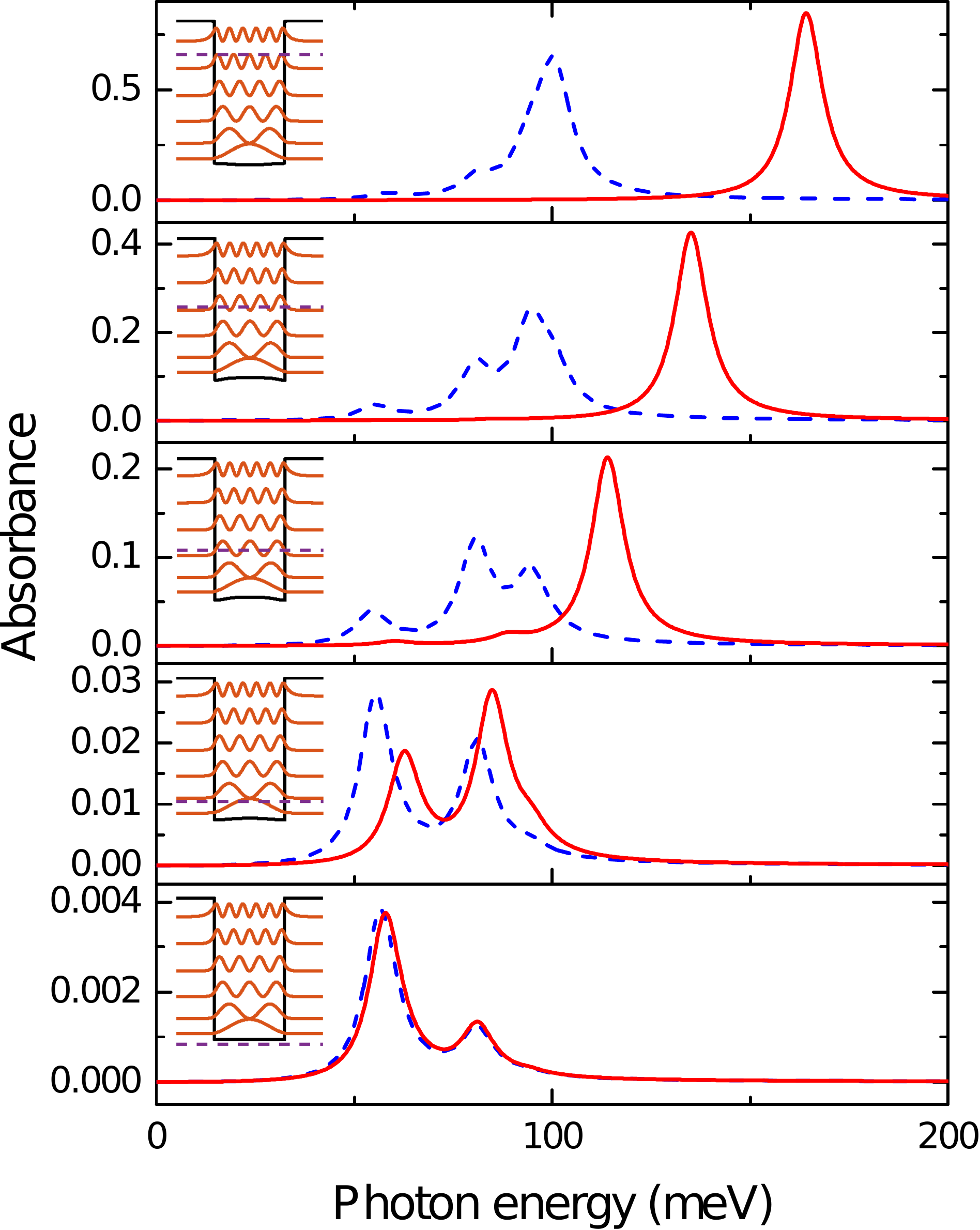}}
\caption{\label{Fig2} Main panels:  The calculated absorption spectra of a $L = 18.5$ nm quantum well with different 2D electron densities $N_{2D}$. From bottom to top: $N_{2D} = 1\times {10}^{11}$ cm$^{-2}$, $1\times {10}^{12}$ cm$^{-2}$, $5\times {10}^{12}$ cm$^{-2}$, $1\times {10}^{13}$ cm$^{-2}$ and $2.2\times {10}^{13}$ cm$^{-2}$. The phenomenological broadening of transitions (full width at half maximum) is 10 meV. The temperature is 300 K. Red continuous lines are the absorption spectra, calculated with Eq.~\eqref{GrindEQ__24_} taking into account Hartree modification of energies and Coulomb coupling of oscillations at different intersubband transitions described by Eq.~\eqref{GrindEQ__8_}. Blue dashed lines are the absorption spectra calculated with Eq.~\eqref{GrindEQ__17_} obtained without taking into account Coulomb coupling. The insets present the band structure, Hartree energy levels, and square moduli of the wave functions.  The Fermi energy corresponding to electron densities in each case is indicated by a (violet) dashed line.}
\end{figure}

\begin{figure}
\center{\includegraphics[width=0.7\textwidth]{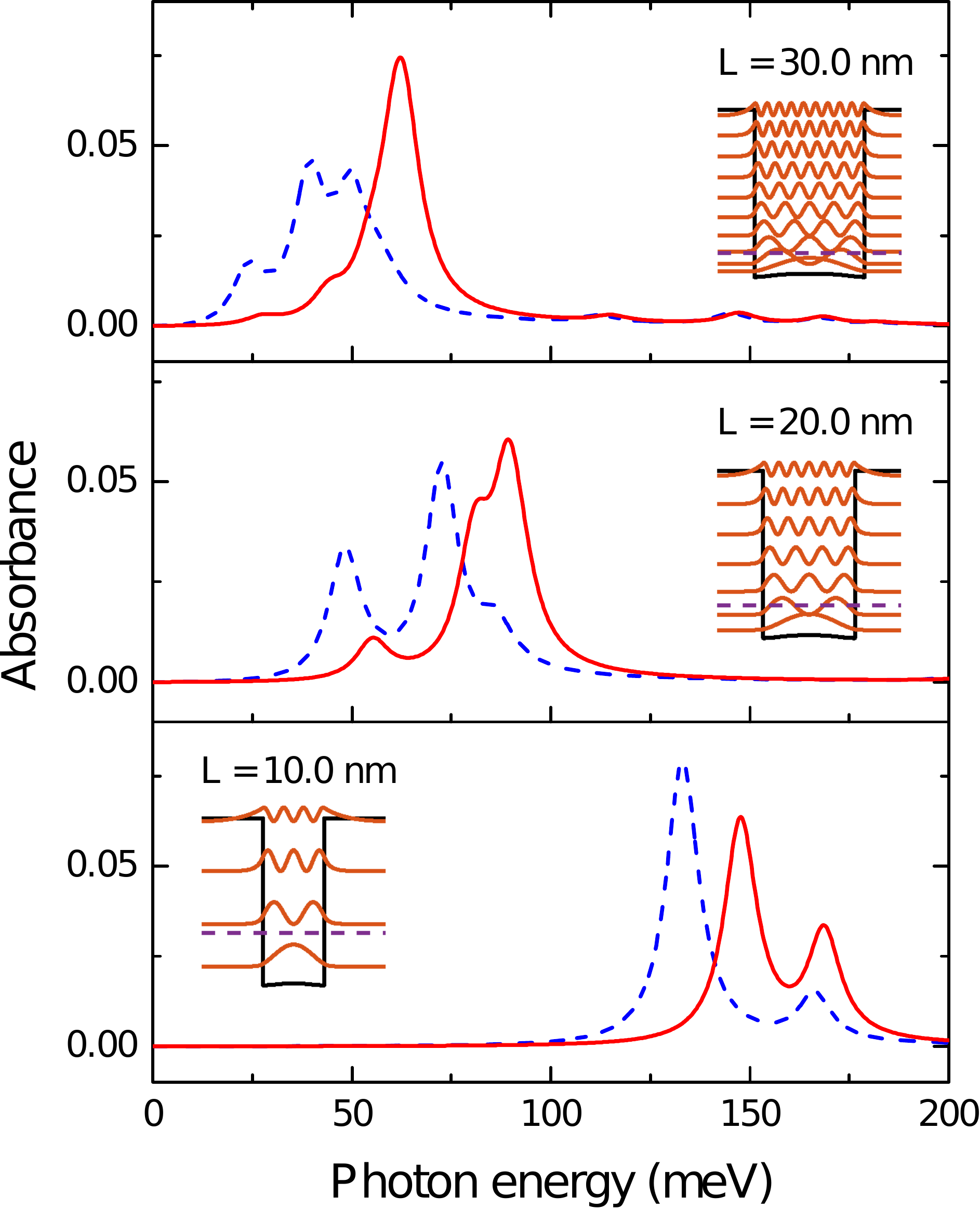}}
\caption{\label{Fig3} The calculated absorption spectra of QWs with different thicknesses; from bottom to top: $L = 10$ nm, 20 nm, and 30 nm, and the same 2D electron density of $2\times {10}^{12}$ cm$^{-2}$. The phenomenological broadening of transitions (full width at half maximum) is 10 meV. The notations are the same as in Fig.~\ref{Fig2}.}
\end{figure}

In the previous sections the analytical estimations for absorption spectra were obtained using rather rough simplifications Eq.~\eqref{GrindEQ__27_}. Surprisingly, these estimations describe the effect quite well, as illustrated by the numerical examples below for a square-well potential. First, we solve numerically Eqs.~\eqref{A1}-\eqref{A4} to find the matrix of transformation from the ``bare'' to Hartree basis, the energies of Hartree levels and their populations. Then we solve for eigenvectors and eigenvalues of $\mathbb{Z}$ matrix in Eq.~\eqref{GrindEQ__10_}. This yields the frequencies of new normal modes $\Omega_l$ and corresponding vectors ${\bm{D}}_{\bm{l}}$ and ${\bm{B}}_{\bm{l}}$ which define the ``oscillator strengths'' of new modes  $\tilde{F}_l$ in Eq.~\eqref{GrindEQ__25_}). The plots of absorption spectra (Eq.~\eqref{GrindEQ__24_}) are presented in Figs.~\ref{Fig2} and \ref{Fig3} for different 2D electron densities in QWs of different thicknesses. The spectra are compared with the ones obtained from Eq.~\eqref{GrindEQ__17_} neglecting the Coulomb coupling. 

The series of plots presented in Fig.~\ref{Fig2} illustrates the transformation of the absorption spectrum with increasing electron density for a fixed quantum well thickness $L=18.5$ nm used in experiments \cite{Delteil12}. At the highest density $N_{2D}=2.2\times {10}^{13}$ cm$^{-2}$ corresponding to the experimental structure in \cite{Delteil12}, five subbands are populated. Using the QW parameters similar to those in \cite{Delteil12}, the resulting calculated absorption spectrum is in a good agreement with the experimental one; compare the top plot in Fig.~\ref{Fig2} with the spectrum in Fig.~2 of \cite{Delteil12}. The analytically predicted strong modification of the spectra is obvious at higher dopings. The frequency of the blueshifted absorption peak agrees with the analytic estimation \eqref{GrindEQ__39_}. It is much higher than the frequencies of the absorption peaks due to ``bare'' intersubband transitions in the model neglecting Coulomb coupling. In fact, the modified spectrum demonstrates almost full transparency at the frequencies of ``bare'' intersubband transitions. The plots in Fig.~\ref{Fig3} show the spectral evolution with varying QW thickness for a electron density.  We see that the larger the width of the well, the greater the relative change in the spectrum associated with the Coulomb coupling. 


\section{\label{Influence}The influence of exchange interaction effects}

\begin{figure}
\center{\includegraphics[width=0.7\textwidth]{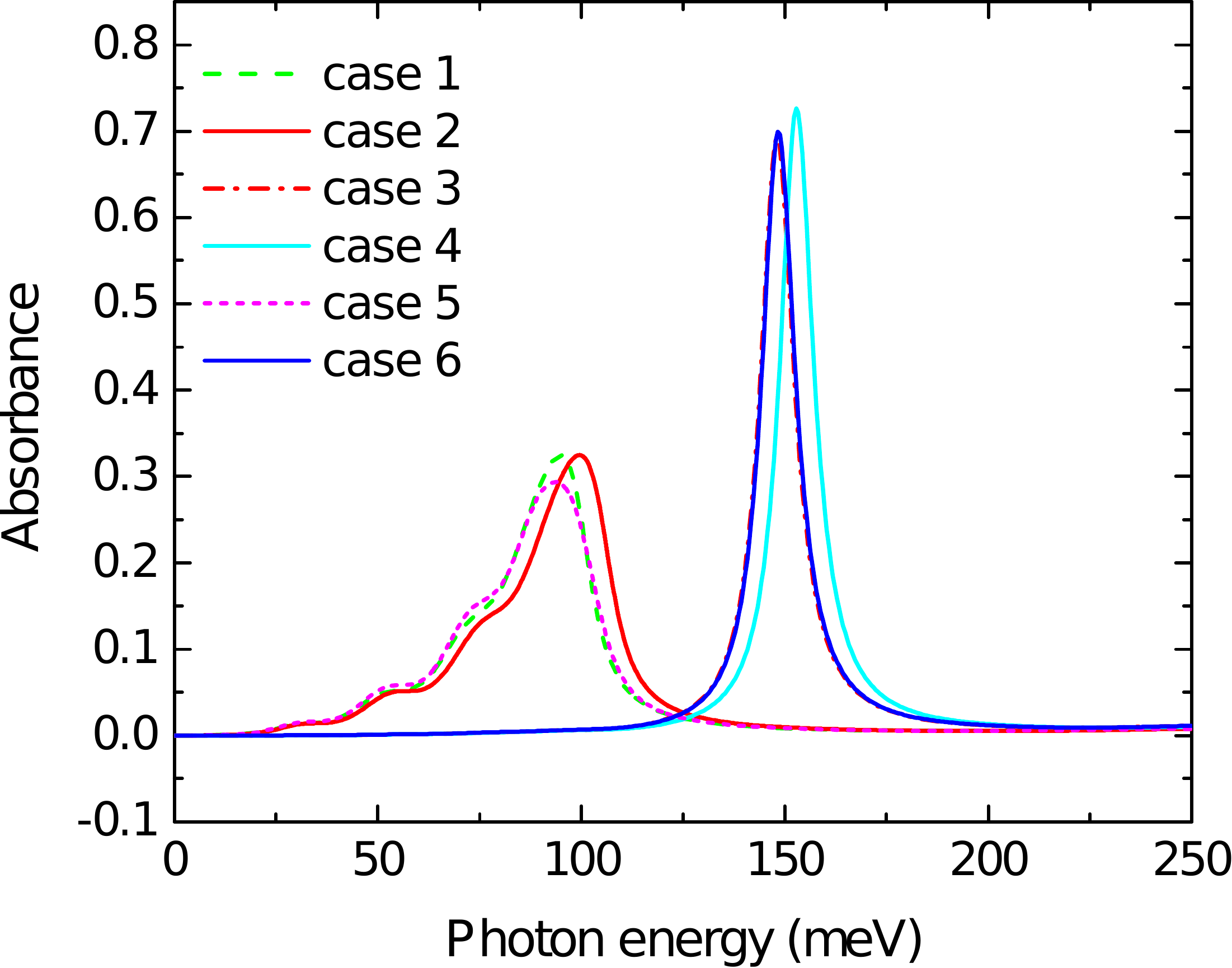}}
\caption{\label{Fig4} The calculated absorption spectra with exchange interaction taken into account perturbatively. The Hartree ground state is treated as unperturbed. Different perturbing terms are taken into account independently. Case 1: without any Coulomb coupling; case 2: with Fock energy renormalixation; case 3: with Hartree coupling (coincides almost exactly with case 6); case 4: with Hartree coupling and Fock energy renormalization; case 5: with Fock coupling and Fock energy renormalization; case 6: with Hartree-Fock coupling and Fock energy renormalization, i.e., with all effects included. The following parameters are used:  $L=18.5$ nm,  $N_{2D}=2.2\times {10}^{13}$ cm$^{-2}$. The phenomenological broadening of transitions (FWHM) is 10 meV.}
\end{figure}

To evaluate the effects of the exchange interaction, we plot the absorption spectrum for the doping density $2.2\times {10}^{13}$ cm$^{-2}$ and QW thickness $L=18.5$ nm by calculating the density matrix elements from Eqs.~\eqref{GrindEQ__7_} which take into account the exchange terms as perturbation to the Hartree ground state. These equations include different exchange (Fock) terms, namely, those responsible for the frequency shift of intersubband transitions (the third row in Eq.~\eqref{GrindEQ__7_}) and for the coupling of coherences (the fourth row in Eq.~\eqref{GrindEQ__7_}), which can be taken into account independently. In Fig.~\ref{Fig4} we show the absorption spectrum when different terms are included. We can see that the blueshifted strong peak is mainly due to Hartree terms in the coupling. Slight blue shift produced by Fock terms in the energy renormalization is almost completely compensated by red shift caused by Fock terms in coupling. This effect of compensation can be explained by comparing different terms in Eqs.~\eqref{GrindEQ__7_}  taking into account expression \eqref{eq_1star} for the overlap integrals $V^{ee}_{mnlp}\left(q\right)$. It is obvious that both ``Fock'' sums in Eqs.~\eqref{GrindEQ__7_} (the third and the fourth rows) are mostly defined by terms proportional to coefficients $V^{ee}_{mnlp}\left(q\right)$ which tend to infinity with $q$ close to zero.  Such divergences can be avoided by taking into account the screening effect (see Appendix~\ref{app D}), but these terms are still prevailing in the sums. The coefficients  $V^{ee,s}_{mnlp}\left(q\right)$  in these terms have indices $m=n,\ l=p$ and only weakly depend on them, so that they can be approximately written as $V^{ee,s}_{mnlp}\left(q\right)\sim {\left.V^{2D,s}\right|}_{q=0}{\delta }_{mn}{\delta }_{lp}\times \mathcal{F}(q)$, where $V^{2D,s}(q)$ is the two-dimensional Fourier transform of the screened Coulomb potential and  $\mathcal{F}(q)$ is a positive-value dimensionless decaying function with characteristic decay scale equal to $1/L$. Here the superscript $s$ stands for screened. Leaving  only these dominant terms, the part of the overall sum responsible for the Fock energy renormalization is 
\begin{eqnarray}
{\left.i\hbar \frac{d}{dt}{\rho }_{nm}\left(\bm{k}\right)\right|}_{Fock\ energy\ renorm}=\nonumber \\ \nonumber
-\frac{1}{S}\sum_{lp}{\sum_{\bm{q}}{{\rho }_{pp}\left(\bm{k}\bm{+}\bm{q}\right)\left(V^{ee}_{nppl}\left(q\right){\rho }_{lm}\left(\bm{k}\right)-V^{ee}_{lppm}\left(q\right){\rho }_{nl}\left(\bm{k}\right)\right)}}\sim\\ \nonumber
   -\frac{1}{S}{\left.V^{2D,s}\right|}_{q=0}\sum_{\bm{q}}{\mathcal{F}(q)\left({\rho }_{nn}\left(\bm{k}\bm{+}\bm{q}\right)-{\rho }_{mm}\left(\bm{k}\bm{+}\bm{q}\right)\right)}{\rho }_{nm}\left(\bm{k}\right) ;
\end{eqnarray} 
whereas the part of the sum which defines the exchange effects in coupling is given by 
\begin{eqnarray}
\nonumber {\left.i\hbar \frac{d}{dt}{\rho }_{nm}\left(\bm{k}\right)\right|}_{Fock\ coupling}=\\
\nonumber -\frac{1}{S}\sum_{p\neq g}{\sum_{\bm{q}}{V^{ee}_{npgm}\left(q\right){\rho }_{pg}\left(\bm{k}\bm{+}\bm{q}\right)}}\left({\rho }_{mm}\left(\bm{k}\right)-{\rho }_{nn}\left(\bm{k}\right)\right)\sim \\
\nonumber   -\frac{1}{S}{\left.V^{2D,s}\right|}_{q=0}\sum_{\bm{q}}{\mathcal{F}(q){\rho }_{nm}\left(\bm{k}\bm{+}\bm{q}\right)}\left({\rho }_{mm}\left(\bm{k}\right)-{\rho }_{nn}\left(\bm{k}\right)\right).
\end{eqnarray} 

The region of wave vectors $\bm{k}\bm{+}\bm{q}\bm{\in }{\bm{\delta }\bm{k}}_{nm}$, where the population difference \(({\rho }_{nn}(\bm{k}+\bm{q})-{\rho }_{mm}(\bm{k}+\bm{q}))\) is far from zero, coincides  with the region where the coherence ${\rho }_{nm}\left(\bm{k}\bm{+}\bm{q}\right)$ is excited.  Replacing these quantities in this region by their mean values  ${\rho }_{nn}\left(\bm{k}\bm{+}\bm{q}\right)-{\rho }_{mm}\left(\bm{k}\bm{+}\bm{q}\right)\approx {\overline{\rho }}_{nn}\left(\bm{k}\right)-{\overline{\rho }}_{mm}\left(\bm{k}\right)$, ${\rho }_{nm}\left(\bm{k}\bm{+}\bm{q}\right)\approx {\overline{\rho }}_{nm}\left(\bm{k}\right)$, we get the following estimations for the two exchange effects: 
\[{\left.i\hbar \frac{d}{dt}{\rho }_{nm}\left(\bm{k}\right)\right|}_{Fock\ energy\ renorm}\sim -\frac{1}{S}{\rho }_{nm}\left(\bm{k}\right)\left({\overline{\rho }}_{nn}\left(\bm{k}\right)-{\overline{\rho }}_{mm}\left(\bm{k}\right)\right){\left.V^{2D,s}\right|}_{q=0}\sum_{\bm{q}\bm{,}\bm{k}\bm{+}\bm{q}\bm{\in }{\bm{\delta }\bm{k}}_{nm}}{\mathcal{F}(q)}\] 
\begin{equation} \label{GrindEQ__42_} 
{\left.i\hbar \frac{d}{dt}{\rho }_{nm}\left(\bm{k}\right)\right|}_{Fock\ coupling}\sim  \frac{1}{S}{\overline{\rho }}_{nm}\left(\bm{k}\right)\left({\rho }_{nn}\left(\bm{k}\right)-{\rho }_{mm}\left(\bm{k}\right)\right){\left.V^{2D,s}\right|}_{q=0}\sum_{\bm{q},\bm{k}\bm{+}\bm{q}\bm{\in }{\bm{\delta }\bm{k}}_{nm}}{\mathcal{F}(q)} .        
\end{equation} 
One can see from here that the blue frequency shift due to Fock terms in energy renormalization and red shift caused by Fock terms in coupling are of the same magnitude but opposite sign and therefore nearly compensate each other.  Indeed, by order of magnitude $\left|{\rho }_{nn}\left(\bm{k}\right)-{\rho }_{mm}\left(\bm{k}\right)\right|\le 2$ and ${\left.V^{2D,s}\right|}_{q=0}\sim \frac{\pi {\hbar }^2}{m^*}$, where the last expression follows from screening theory presented in the Appendix~\ref{app D}. Furthermore, the number of electron states in the region of wave vectors $\bm{k}\bm{+}\bm{q}\bm{\in }{\bm{\delta }\bm{k}}_{\bm{\ }nm}$ is of the order of  $S\frac{m^*}{2\pi \hbar }{\omega }_{mn}$. Then for the frequency shifts we get
\[{\mathrm{\Delta }}_{Fock\ energy\ renorm}\approx -{\mathrm{\Delta }}_{Fock\ coupling}\sim {\omega }_0\times o\left({\left(k_FL\right)}^{-1}\right), \] 
where in notations of Section~\ref{Coulomb} ${\omega }_0$ is the average transition frequency, $k_F$ is the Fermi wave number for a typical transition, and the small value $o\left({\left(k_FL\right)}^{-1}\right)$ is defined by the decaying function $\mathcal{F}\left(q\right)$. Each of these frequency shifts separately is smaller than $\omega_0$ and their difference is even much smaller.  The result is confirmed by numerical calculations in Fig.~\ref{Fig4} and provides the rationale for neglecting the exchange terms when calculating the absorption spectra of highly doped QWs. 


\section{\label{Discussion}Conclusions}

We presented a consistent theoretical explanation of the effect of Coulomb-induced collapse of multiple intersubband absorption peaks in highly doped quantum wells into one strong and extremely blueshifted peak. The theory is based on the density matrix equations taking into account pairwise Coulomb interactions of electrons within the Hartree-Fock (HF) approximation. We show that in the high-doping limit the optical response is described by linearly coupled $2N$ first-order differential equations for intersubband coherences, where $N$ is the total number of the intersubband transitions.  Therefore, the observed spectra can be understood within an intuitive and transparent picture of self-synchronization in a system of $N$ coupled oscillators, which has numerous analogies including the exact mechanical analogy. 
Analytic expressions are obtained for the frequencies and oscillator strengths of the new collective normal modes of the system, renormalized by strong Coulomb interaction through the collective field.  In the high doping regime, Coulomb-induced synchronization leads to a merger of all intersubband absorption resonances into one sharp peak at the frequency substantially higher than all ``bare'' intersubband transitions and accumulating all their oscillator strength.

\appendix
\section{\label{app A} The Hartree basis}

The eigen functions of the Hartree Hamiltonian Eq.~\eqref{GrindEQ__5_} can be found from equation Eq.~\eqref{GrindEQ__6_} written in ``bare'' (single-particle) basis, using the expansion over the eigen functions of a single-particle Hamiltonian ${\hat{H}}^0$,  ${\varphi }_m\left(z\right)=\sum_n{c^n_m}{\varphi }^0_n(z)$:
\begin{equation}
\label{A1}
\sum_m{H^H_{nm}}c^m_l=E^H_lc^n_l,
\end{equation}         
where the matrix elements of the Hartree Hamiltonian in the ``bare basis'' are given by
\begin{equation}
\label{A2}
H^H_{nm}=E^0_n{\delta }_{nm}-N_{2D}{\tilde{V}}^{ei}_{nm}+\sum_{lpq}{{\rho }_{ll}{\left(c^p_l\right)}^*c^q_l{\tilde{V}}^{ee}_{nmpq}}.     
\end{equation}
Here the overlap integrals ${\tilde{V}}^{ei}_{nm}$ and ${\tilde{V}}^{ee}_{nmpq}$ given by Eq.~\eqref{GrindEQ__3_} and Eq.~\eqref{GrindEQ__4_} are calculated over ``bare'' basic functions. ${\rho }_{ll}=\frac{1}{S}\sum_{\bm{k}}{{\rho }_{ll}\left(\bm{k}\right)}$ is the Hartree subband population, which in the case of an equilibrium Fermi distribution should be self-consistently calculated as
\begin{equation}
\label{A3}
{\rho }_{ll}=\frac{m^*}{\pi {\hbar }^2}\left(E_F-E^H_l\right),
\end{equation}
where $E_F$ is Fermi energy and  
\begin{equation}
\label{A4}
N_{2D}=\sum_l{{\rho }_{ll}}.
\end{equation} 

Equations \eqref{A1}-\eqref{A4} can be used for numerical calculation of Hartree states in the case of known ``bare'' states.

\section{\label{app B} The sum rule for new collective normal modes}

Here we prove that the sum of new ``oscillator strengths'' defined for the normal modes is equal to the sum of ``oscillator strengths'' in the system of uncoupled partial oscillators:
\begin{equation}
\label{B1}
\sum^{2N_t}_{l=1}{{\tilde{F}}_l}=\sum^{2N_t}_{j=1}{F_j}=1.        
\end{equation}
Taking into account Eq.~\eqref{GrindEQ__25_} we can rewrite the first sum in the form 
\[\sum^{2N_t}_{l=1}{{\tilde{F}}_l}=\frac{2m^*}{e^2N_{2D}E^{\omega }}\sum^{2N_t}_{l=1}{{\mathrm{\Omega }}_l\tilde{{\mu }_l}{\tilde{f}}^{\omega }_l}=\frac{2m^*}{e^2N_{2D}E^{\omega }}\sum^{2N_t}_{l,j}{D_{jl}{\mu }_j\sum^{2N_t}_{k=1}{B_{lk}f^{\omega }_k{\mathrm{\Omega }}_l}}\] 
Then, using the relation $D_{jl}{\mathrm{\Omega }}_l=\sum^{2N_t}_i{Z_{ji}D_{il}}$ we obtain
\[\sum^{2N_t}_{l=1}{{\tilde{F}}_l}=\frac{2m^*}{e^2N_{2D}E^{\omega }}\sum^{2N_t}_{ljik}{{\mu }_jB_{lk}f^{\omega }_kZ_{ji}D_{il}}\] 
Taking into account the relation $D_{ij}={\left(B^{-1}\right)}_{ij}$ and Eq.~\eqref{GrindEQ__10_}, the following transformation is possible:
\[\sum^{2N_t}_{l=1}{{\tilde{F}}_l}=\frac{2m^*}{e^2N_{2D}E^{\omega }}\sum^{2N_t}_{jik}{{\mu }_jf^{\omega }_kZ_{ji}{\delta }_{ki}}=\frac{2m^*}{e^2N_{2D}E^{\omega }}\sum^{2N_t}_k{{\mu }_kf^{\omega }_k{\omega }_k}+\frac{2m^*}{\hbar N_{2D}E^{\omega }}\sum^{2N_t}_{jk}{{\mu }_jf^{\omega }_kI_{jk{'}}\mathrm{\Delta }n_j}.\] 
Using Eq.~\eqref{GrindEQ__13_} and the natural assumption that one-dimensional basic functions are real, so that   $I_{jk{'}}=I_{jk}$,  it can be shown that the second sum is equal to zero: 
\[\frac{2m^*}{\hbar N_{2D}E^{\omega }}\sum^{2N_t}_{jk}{{\mu }_jf^{\omega }_kI_{jk{'}}\mathrm{\Delta }n_j}=\frac{2m^*}{{\hbar }^2N_{2D}}\sum^{2N_t}_{jk}{{\mu }_j{\mu }_k\mathrm{\Delta }n_j\mathrm{\Delta }n_kI_{jk}}=\] 
\[\frac{2m^*}{{\hbar }^2N_{2D}}\sum^{2N_t}_j{{\mu }_j\mathrm{\Delta }n_j\sum^{N_t}_{k,{\omega }_k>0\ }{{\mu }_k(\mathrm{\Delta }n_kI_{jk}-\mathrm{\Delta }n_kI_{jk})}}=0. \] 
As result, taking into account Eq.~\eqref{GrindEQ__18_}, we obtain Eq.~\eqref{B1}.

\section{\label{app C} The mechanical model of ``Coulomb springs''}

Here write the equations of motion for the model of mechanical oscillators shown in Fig.~\ref{Fig1}. We consider one-dimensional (along axis $x$) mirror-symmetric oscillation mode of ``upper'' (blue) and ``lower'' (yellow) masses $m$ coupled in pairs by green springs, when their coordinates are equal in absolute value and have a different sign. Let $N_t$ be a total number of such oscillators (in Fig.~\ref{Fig1} $N_t=3$). The grey ``Coulomb springs'' couple each ``upper'' mass with all ``lower'' masses and vice versa. The equations for the distances $l_i$ between the two masses in each oscillator are
\[\frac{1}{2}m\frac{d^2}{{dt}^2}\mathrm{\Delta }l_i=-k_i{\mathrm{\Delta }l}_i-K{\mathrm{\Delta }l}_i-\frac{1}{2}\sum_{j\neq i}{K\left({\mathrm{\Delta }l}_i+{\mathrm{\Delta }l}_j\right)},\] 
where ${\mathrm{\Delta }l}_i=l_i-l_0$,  $l_0$ is the equilibrium spring length, $k_i$ are spring constants of green springs which define partial frequencies of oscillations ${\omega }_i=\sqrt{\frac{2k_i}{m}}$. Note that the ``Coulomb springs'' not only couple different oscillators but also strengthen the coupling between the masses within each oscillator. Under the condition of a large stiffness of ``Coulomb springs'' as compared with the stiffness of green springs, 
\[KN_t\gg k_i\] 
there is obviously a normal mode with in-phase motion of all oscillators  ${\mathrm{\Delta }l}_i$$=\mathrm{\Delta }l$:
\[\frac{d^2}{{dt}^2}\mathrm{\Delta }l=-{\mathrm{\Omega }}^2_m\mathrm{\Delta }l.\] 
Its frequency is defined by the relation  
\[{\mathrm{\Omega }}_m\approx \sqrt{2\frac{K}{m}N_t}.\] 
It is analogous to Eq.~\eqref{GrindEQ__31_}.

\section{\label{app D} The screening effect of higher-order correlations}

Beyond the Hartree-Fock approximation, the screening effect coming from higher order correlations should be included. To do this we need to replace the overlap integrals of the Coulomb potential $V^{ee}_{mnlp}\left(q\right)$ in Eq.~\eqref{eq_1star} with their screened values $V^{ee,s}_{mnlp}\left(q\right)$ in accordance with the following relations \cite{Sotirelis93}:
\[V^{ee}_{mnlp}\left(q\right)=\sum_{ij}{{\epsilon }_{mjni}\left(\bm{q}\right)V^{ee,s}_{ijlp}\left(q\right)} ,\] 
where
\[{\epsilon }_{mjni}\left(\bm{q}\right)={\delta }_{mi}{\delta }_{nj}-V^{ee}_{mnji}\left(q\right){\mathrm{\Pi }}^0_{mn}\left(\bm{q}\right),\] 
with
\begin{equation}
\label{D1}
{\mathrm{\Pi }}^0_{mn}\left(\bm{q}\right)={\mathop{\mathrm{lim}}_{\varepsilon \to 0^+} \sum_{\bm{k}}{\frac{{\rho }_{mm}\left(\bm{k}\bm{+}\bm{q}\right)-{\rho }_{nn}\left(\bm{k}\right)}{\left(H^0_{mm}\left(\bm{k}\bm{+}\bm{q}\right)-H^0_{nn}\left(\bm{k}\right)\right)-i\varepsilon }}\ } .      
\end{equation}
Note that we use static screening \cite{Li03}, \cite{Haug04}.

When the denominator in Eq.~\eqref{D1} is zero, the numerator will also be zero, so we need to consider the ways these two limits are approached. If $\left|\bm{q}\right|\to 0$ and $m=n$, and we have
\[{\mathrm{\Pi }}^0_{nn}\left(\bm{q}\to 0\right)=-\frac{m^*}{2\pi {\hbar }^2}{\rho }_{nn}\left(\bm{k}=0\right).\] 
If $\left|\bm{q}\right|\neq 0$, the denominator in Eq.~\eqref{D1} can still be zero, let us say at $\bm{k}={\bm{k}}_{\bm{0}}$, in which case  
\[\frac{{\rho }_{mm}\left({\bm{k}}_{\bm{0}}\bm{+}\bm{q}\right)-{\rho }_{nn}\left({\bm{k}}_{\bm{0}}\right)}{\left(H^0_{mm}\left({\bm{k}}_{\bm{0}}\bm{+}\bm{q}\right)-H^0_{nn}\left({\bm{k}}_{\bm{0}}\right)\right)}={\left.\frac{\partial {\rho }_{nn}(\bm{k})}{\partial H^0_{nn}(\bm{k})}\right|}_{\bm{k}\bm{=}{\bm{k}}_{\bm{0}}}={\left.\frac{\partial {\rho }_{mm}(\bm{k})}{\partial H^0_{mm}(\bm{k})}\right|}_{\bm{k}\bm{=}{\bm{k}}_{\bm{0}}\bm{+}\bm{q}}.\] 
So, the elements in the summation of Eq.~\eqref{D1} are always well-defined.

\bibliography{QW_Coulomb.bib}

\end{document}